\documentclass{aa}

\usepackage{graphicx}
\usepackage{txfonts}
\usepackage{lipsum}
\usepackage{subcaption}
\usepackage{lscape}             
\usepackage{placeins}
\usepackage{siunitx}
\DeclareSIUnit \parsec {pc} 
\DeclareSIUnit \Msol {M_{\odot}} 
\DeclareSIUnit \Lsol {L_{\odot}} 
\usepackage{float} 
\usepackage{booktabs}

\usepackage{natbib,twoopt}

\usepackage[breaklinks=true]{hyperref}

\begin{document}
	\title{The p-Laplacian as a Framework for Generalizing Newtonian Gravity and MoND}
	
	\subtitle{}
	
	\author{D. Scherer\inst{1}
		\and J. Pflamm-Altenburg\inst{2} \and P. Kroupa\inst{2}${}^,$\inst{3}\and E. Gjergo\inst{4}${}^,$\inst{5} 
	}
	
	\institute{Universität Bonn, 53115 Bonn, Germany\\
		\email{s6dvsche@uni-bonn.de}
		\and Helmholtz-Institut für Strahlen- und Kernphysik, Universität Bonn, Nussallee 14-16, 53115 Bonn, Germany
		\and Astronomical Institute, Charles University, V Holesovickach 2, 18000 Praha, Czech Republic
		\and School of Astronomy and Space Science, Nanjing University, Nanjing 210093, People's Republic of China 
		\and Key Laboratory of Modern Astronomy and Astrophysics (Nanjing University), Ministry of Education\\ }
	
	\date{Received March 27 2025 / Accepted April 21 2025}
	
	\abstract
	{The Radial Acceleration Relation (RAR) follows from Milgromian gravitation (MoND) and velocity dispersion data of many dwarf spheroidal galaxies (dSphs) and galaxy clusters have been reported to be in tension with it.}
	{We consider the Generalized Poisson Equation (GPE), expressed in terms of the p-Laplacian, which has been applied in electrodynamics, and investigate whether it can address these tensions.} 
	{From the GPE we derive a generalized RAR characterized by the $p$-parameter from the p-Laplacian and a velocity dispersion formula for a Plummer model. We apply these models to Milky Way and Andromeda dSphs and HIFLUGS galaxy clusters and derive a $p$-parameter for each dSph and galaxy cluster. We explore a relation of $p$ to the mass density of the bound system, and alternatively a relation of $p$ to the external field predicted from Newtonian gravity}
	{This ansatz allows the deviations of dSphs and galaxy clusters from the RAR without the need of introducing dark matter. Data points deviate from the Milgromian case, $p=3$, with up to $5\sigma$-confidence. Also, we find the model predicts velocity dispersions, each of which lies in the 1$\sigma$-range of their corresponding data point allowing the velocity dispersion to be predicted for early-type dwarf satellite galaxies from their baryonic density. The functional relation between the mass density of the bound system and $p$ suggests $p$ to increase with decreasing density. We find for the critical cosmological density  $p(\rho_{\text{crit}}) = \num{12.27(39)}$. This implies significantly different behaviour of gravitation on cosmological scales. Alternatively, the functional relation between $p$ and the external Newtonian gravitational field suggests $p$ to decrease with increasing field strength.}
	{The GPE fits the RAR data of dSphs and galaxy clusters, reproduces the velocity dispersions of the dSphs, gives a prediction for the velocity dispersion of galaxy clusters from their baryonic density and may explain the non-linear behaviour of galaxies in regions beyond the Newtonian regime.}
	
	
	\maketitle

	\section{Introduction}
	\label{sec:intro}
	
	Modified Newtonian Dynamics (MoND) was introduced by \citet{1983ApJ...270..365M}. It is a non-relativistic modification of the Newtonian law of gravity, which can largely explain the dynamics in galaxies without the need of introducing dark matter. This constitutes an important research program \citep{Merritt_2020} given the tensions between observations and dark-matter-based models \citep{kroupa2023tensionsdarkmatterbasedmodels}. Still, the acceleration $a_0$ known as Milgrom's constant needs to be introduced in the theory and must be obtained from data. The value $a_0 = \qty[exponent-product=\cdot]{1.20(2)E-10}{\metre\per\second\squared}$ \citep{2017ApJ...836..152L} is used for data analysis in Chapter \ref{sec:ana}. MoND replicates Newtonian behaviour for large gravitational accelerations $ a \gg a_0$, while postulating a logarithmic potential for $a \ll a_0$, recently verified through weak lensing by \citet{Mistele_2024}.  If the gravitational acceleration as predicted by MoND is plotted against the gravitational acceleration as predicted by Newton's law of gravity $g_{\text{N}}$, a deviation from the line of identity for $a < a_0$ is obtained, so that $a > g_{\text{N}}$. The functional relation $a(g_{\text{N}})$ is called the Radial Acceleration Relation (RAR). {The velocity dispersions of many dwarf spheroidal satellite galaxies (dSphs) and galaxy clusters have been reported to be in tension with the RAR \citep{Safarzadeh_2021, 2017ApJ...836..152L, Sanchez-Salcedo_2007, 2008MNRAS.387.1481A, 2013ApJ...766...22M, 2013ApJ...775..139M, Alexander_2017}. This tension can be interpreted to imply that MoND might not be a generally valid effective description of non-relativistic gravitation. 
		However, this tension needs to be seen in view of the multiple flavours of dSphs among the data of \citet{2017ApJ...836..152L} and  \citet{Safarzadeh_2021}: those that lie along the RAR, those that should not be on the RAR in MoND due to the external field effect (EFE, as predicted by MoND), those that are subject to tidal disruption (a relevant process also in Newtonian dynamics as shown by \citet{1997NewA....2..139K} and applied to dSph data by \citet{2010ApJ...722..248M}), and those that might be genuinely offset. The dSphs that fall on the RAR generally (though not always) have the best data and those that are offset have consistently less-reliable data.  In MoND, one must distinguish between isolated dSphs and those subject to the EFE. The former should follow the RAR but the latter should not. Both of these situations assume dynamical equilibrium, but there is a narrow window in which that holds before the EFE becomes tidally disruptive. Judging from the simulations by \citet{2000ApJ...541..556B}, some dSphs should be out of equilibrium and off the RAR \citep{1997NewA....2..139K}. 
		The complexity of the gravitational stellar-dynamics allows for new ansatzes \citep{2021MNRAS.507..803S, 2021AJ....162..202M}. Here, we show that by generalizing gravitation in the regime $a < a_0$ with the help of the p-Laplacian permits a comprehensive accommodation of dSphs and galaxy clusters.}
	
	The Generalized Poisson Equation (GPE) reads \citep{EVANS1982356}:
	\begin{equation}
		\nabla \cdot \left(\left(\frac{\left|\nabla\Phi(\vec{x})\right|}{a_0}\right)^{p-2} \nabla \Phi(\vec{x})\right) = 4\pi G \rho(\vec{x})\;.
		\label{eq:plapl}
	\end{equation}
	$\Phi(\vec{x})$ is the gravitational potential, such that the acceleration $\vec{a}(\vec{x}) = - \nabla \Phi$ with $a = |\vec{a}|$. $G$ is the gravitational constant, $\rho(\vec{x})$ the mass density distribution, and $\vec{x}$ the position vector. $p$ is a real dimensionless parameter with $p > 1$ \citep{LindqvistpLaplace}. For $p=2$ follows the Newtonian potential, for $p=3$ the logarithmic Milgromian potential is obtained outside of $\rho(\vec{x})$. Aside of these two $p$ values, there is no physical meaning of $p$ known in the context of gravitational theory. The data analyzed here allow the possibility that the p-Laplacian approach combines the non-equilibrium dynamics (EFE and tidal disruption) through a varying p-value. In electrodynamics the p-Laplacian with arbitrary $p$ is used to describe electrorheological fluids, which are a special sort of non-linear dielectrics \citep{alma991043454719706467}. There, $p$ is a material constant and can be a function of the electric field. Interpreting $p$ as the dimension of three-dimensional or higher-dimensional Euclidean space, \citet{Milgrom_1997} showed the GPE to be conformally invariant. The analysis presented here does not follow his interpretation, but investigates a functional relation between $p$ and $\rho(\vec{x})$ assuming $\rho(\vec{x})$ to be isolated, and a possible functional relation between $p$ and an external Newtonian gravitational field $\vec{g}_\text{tot}$. This analysis considers simple solutions for arbitrary $p$ in spherical symmetry in Sect. \ref{sec:deriv}, applies them to data on dSphs and galaxy clusters in Sect. \ref{sec:ana}, elaborates on MoND being  possibly not fundamental dynamics in Sect. \ref{sec:ext}, and discusses possible physical implications in Sect. \ref{sec:dis}.
	
	\section{Derivation of the RAR, gravitational potential and velocity dispersion assuming Plummer models for arbitrary $p$}
	\label{sec:deriv}
	\subsection{RAR}
\begin{figure}[h!] 
		\centering
		\includegraphics[width=0.5\textwidth]{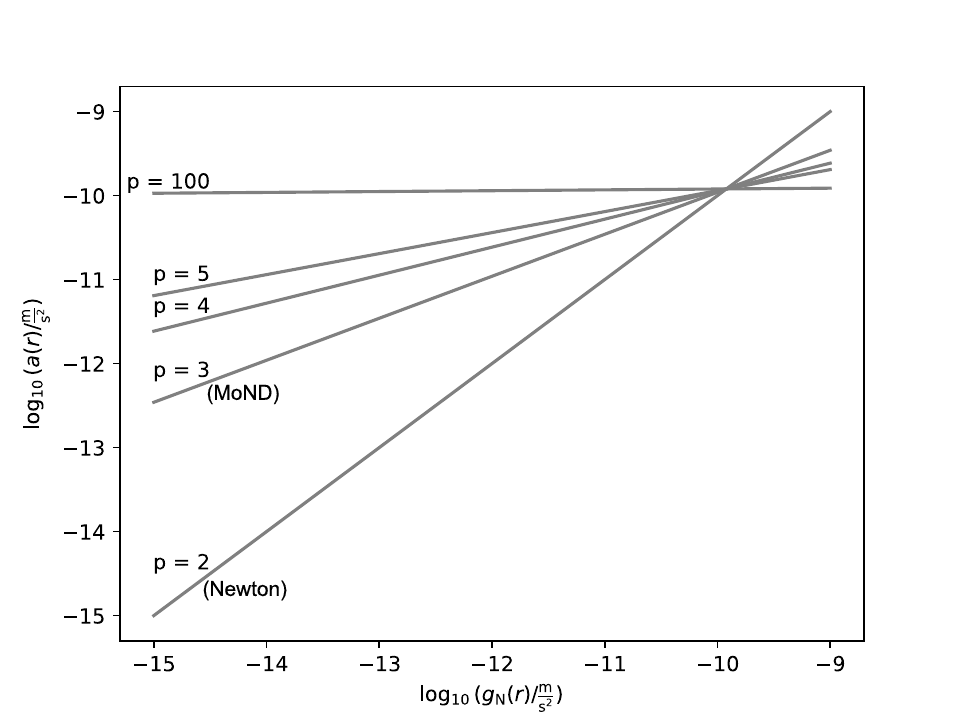}
		\caption{RAR for arbitrary $p$. The bold grey lines indicate the RAR for different $p$ (Eq. \eqref{eq:prar}).}
		\label{fig:1}
	\end{figure}

	In this subsection, the RAR is derived for arbitrary $p$ from the GPE, Eq. \eqref{eq:plapl}, as a static spherically symmetric problem. This generalized RAR enables to obtain the $p$-parameters in Sect. \ref{sec:ana}. The GPE simplifies in spherical symmetry with $a(r):=\left|\partial_r\Phi\right|$ to the expression:
	\begin{equation}
		\label{eq:plaplsph}
		\frac{1}{r^2}\partial_r\left(r^2\left(\frac{a(r)}{a_0}\right)^{p-2} a(r)\right) = 4\pi G \rho(r)\;.
	\end{equation}
	The Newtonian acceleration $g_\text{N}(r)$ at distance $r$ in a spherical volume $V=\frac{4\pi}{3}r^3$ is given by
	\begin{align}
		g_{\text{N}}(r) &= \frac{GM(r)}{r^2} = \frac{4\pi G}{r^2} \int\limits_{0}^{r} \rho(r') r'^2\, \text{d}r'\;, \; \text{with the mass} \\
		M(r)& := 4 \pi \int\limits_{0}^{r} \rho(r') r'^2\, \text{d}r'\;.
	\end{align}	
	Solving the GPE, Eq. \eqref{eq:plaplsph} by radial integration leads to the $p$-dependent relation between the fundamental and Newtonian accelerations:
	\begin{equation}
		\frac{a(r)}{a_0} = \sqrt[p-1]{\frac{g_{\text{N}}(r)}{a_0}}\;, \label{eq:prar}
	\end{equation}
	and is shown in Fig. \ref{fig:1}. 
	In the Newtonian case $p=2$, the RAR becomes the identity. For large $p$:
	\begin{equation}
		\underset{p \to \infty} \lim  a(r)= a_0 \label{eq:ainf} \;.
	\end{equation}
	Solving Eq. \eqref{eq:prar} for $a(r)$ yields:
	\begin{equation}
		a(r) = \sqrt[p-1]{\frac{GM(r)a_0^{p-2}}{r^2}}\;. \label{eq:acc}
	\end{equation}
	In the cases $p=2$ and $p=3$, Eq. \eqref{eq:acc} simplifies to the Newtonian and Milgromian gravitational accelerations:
	\begin{align}
		p=2 &\Rightarrow a(r) = \frac{GM(r)}{r^2}\;, \\
		p=3 &\Rightarrow a(r) = \frac{\sqrt{GM(r)a_0}}{r}\;.
	\end{align}
	
	\subsection{The velocity dispersion of the generalized Plummer mass distribution} 
	In this subsection, the velocity dispersion for arbitrary $p$ and a Plummer model is derived. The Plummer model is a simple and proper density distribution to describe dSphs. The velocity dispersion cannot be calculated independently of the density distribution, if $p \neq 3$. For a proof, see Appendix \ref{sec:app2}.  Consider the gravitational acceleration for arbitrary $p$ (Eq. \eqref{eq:acc}) and the Plummer model \citep[p. 81 in ][]{2003gmbp.book.....H}:
	\begin{align}
		M(r) &= M_\text{st}\frac{\left(\frac{r}{r_\text{pl}}\right)^3}{\left(1+\left(\frac{r}{r_\text{pl}}\right)^2\right)^{\frac{3}{2}}}\;, \label{eq:plummas} \\ \rho(r) &= \frac{3M_\text{st}}{4\pi r_\text{pl}^3} \frac{1}{\left(1+\left(\frac{r}{r_\text{pl}}\right)^2\right)^{\frac{5}{2}}} \label{eq:plumden}\;,
	\end{align}
	with Plummer radius $r_\text{pl}$ and stellar mass $M_\text{st}$.
	The radial component of the isotropic velocity dispersion, $\sigma_r^2(r)$, obtained from the Jeans equation, reads \citep[Chap. 4.8 in ][]{2008gady.book.....B}:
	\begin{equation}
		\sigma_r^2(r) = \frac{1}{\rho(r)}\int_r^\infty\rho(r')a(r')\text{d}r' \label{eq:veldisp} \;.
	\end{equation}
	Inserting Eqs. \ref{eq:acc}, \ref{eq:plummas} and \ref{eq:plumden} into Eq. \eqref{eq:veldisp} yields:
	\begin{align}
		\sigma_r^2(r) &= C(p) \big(1+\big(\frac{r}{r_\text{pl}}\big)^2\big)^{\frac{5}{2}} \int\limits_r^\infty \frac{1}{\big(1+\big(\frac{r'}{r_\text{pl}}\big)^2\big)^{\frac{5}{2}}} \Bigg(\frac{\frac{r'}{r_\text{pl}}}{\big(1+\big(\frac{r'}{r_\text{pl}}\big)^2\big)^\frac{3}{2}}\Bigg)^\frac{1}{p-1} r_\text{pl}^\frac{-2}{p-1} \text{d}r'\\
		&= C(p) \big(1+\big(\frac{r}{r_\text{pl}}\big)^2\big)^{\frac{5}{2}} r_\text{pl}^\frac{p-3}{p-1} \int\limits_{q(r)}^\infty \Bigg(\frac{q(r')}{\big(1+q(r')^2\big)^\frac{5p-2}{2}}\Bigg)^\frac{1}{p-1}\text{d}q(r') \\
		&= \frac{p-1}{p} C(p) \big(1+\big(\frac{r}{r_\text{pl}}\big)^2\big)^{\frac{5}{2}} r_\text{pl}^\frac{p-3}{p-1} \Big( \lim_{q(r) \rightarrow \infty}I(q(r),p)- I(q(r),p)\Big)\;, \label{eq:sigmar}
	\end{align}
	after introducing the following definitions:
	\begin{align}
		q(r)&:= \frac{r}{r_\text{pl}}\;, \\
		C(p) &:=\sqrt[p-1]{GM_\text{st}a_0^{p-2}}\;, \\
		I(q,p) &:= q^\frac{p}{p-1} {}_2F_1\Big(\frac{2-5p}{2-2p}, \frac{p}{2p-2};\frac{3p-2}{2p-2};-q^2\Big)\;,
	\end{align}
	where ${}_2F_1(a,b;c;z)$ is the hypergeometric function.
	
	The characteristic 1D velocity dispersion $\sigma_\text{1D}^2 := \frac{2}{3} \frac{K}{M_\text{st}}$ is obtained. $K:= \frac{3}{2}\int\limits_0^\infty \sigma_r^2(r) \rho(r) 4 \pi r^2\, \text{d}r$ is the kinetic energy \citep[Chap. 4.8 in ][]{2008gady.book.....B} for the Plummer sphere.
	\begin{equation}
		\sigma_\text{1D}^2 = \frac{1}{M_\text{st}} \int\limits_0^\infty \sigma_r^2(r) \rho(r) 4 \pi r^2\, \text{d}r\;.
	\end{equation} 
	Inserting Eq. \eqref{eq:sigmar} for $\sigma_r(r)$ and Eq. \eqref{eq:plumden} for $\rho(r)$ yields:
	\begin{equation}
		\sigma_\text{1D}^2 = 3 \Omega(p) r_\text{pl}^{-3} \cdot \int\limits_0^\infty \Lambda(q(r),p) r^2 \, \text{d}r \;, \label{eq:sigmaplum}
	\end{equation}
	with 
	\begin{align}
		\Omega(p) &:= \frac{p-1}{p}\cdot C(p)\cdot r_\text{pl}^{\frac{p-3}{p-1}} \;, \\
		\Lambda(q(r),p) &:= \Big( \lim_{q(r) \rightarrow \infty}I(q(r),p)- I(q(r),p)\Big) \;.
	\end{align}
	Converting the integration variable to $q(r)$ and integrating by parts results in:
	\begin{align}
		\sigma_\text{1D}^2 &= \Omega(p) \cdot \Big(  \big[q(r)^3 \Lambda(q(r),p)\big]_{q(r)=0}^{q(r) \rightarrow \infty} \\
		&- \int\limits_0^\infty q(r)^3 \partial_{q(r)}\Lambda(q(r),p) \, \text{d}q(r) \Big) \\
		&= \Omega(p) \cdot \Big(  \big[q(r)^3 \Lambda(q(r),p)\big]_{q(r)=0}^{q(r) \rightarrow \infty} \\ &+ \frac{p}{p-1} \cdot \int\limits_0^\infty q(r)^3 \Bigg(\frac{q(r)}{\big(1+q(r)^2\big)^\frac{5p-2}{2}}\Bigg)^\frac{1}{p-1} \, \text{d}q(r) \Big) \\
		&= \Omega(p) \cdot \lim_{q\rightarrow \infty}\Bigg( q^3 \Lambda(q,p) + I'(q,p) \Bigg) \;, 
	\end{align}
	{with}
	\begin{equation}
		I'(q,p) := \frac{p}{4p-3} q^{\frac{4p-3}{p-1}}{}_2F_1\Big(\frac{2-5p}{2-2p}, \frac{3-4p}{2-2p}; \frac{5-6p}{2-2p}; -q^2 \Big) \;.
	\end{equation}
	In the last step, the lower limit $q(r)$ of the integral (Eq. \eqref{eq:sigmaplum}) does not contribute: For $p>1$, $\lim_{q\rightarrow 0}(\Lambda(q,p))$ is finite, $\lim_{q\rightarrow 0}(q^3) = 0$, $\lim_{q\rightarrow 0}(q^\frac{4p-3}{p-1}) = 0$ and ${}_2F_1\Big(\frac{2-5p}{2-2p}, \frac{3-4p}{2-2p}; \frac{5-6p}{2-2p}; 0 \Big) =1$.
	Using l'H$\hat{\text{o}}$spital's rule for
	\begin{equation}
		\lim_{q\rightarrow \infty}\frac{\Lambda(q,p)}{q^{-3}} = \lim_{q\rightarrow \infty}\frac{\partial_q \Lambda(q,p)}{-3q^{-4}} = \lim_{q\rightarrow \infty} q^\frac{-5p-1}{p-1} = 0\;, \text{if}\; p>1 \;,
	\end{equation}
	leads to the final expression for $\sigma_\text{1D}^2$:
	\begin{equation}
		\sigma_\text{1D}^2 = \Omega(p) \cdot \lim_{q \rightarrow \infty} I'(q,p) \label{eq:sigmafin} \;.
	\end{equation}
	In Sect. \ref{sec:ana}, velocity dispersions for dSphs are calculated with Eq. \eqref{eq:sigmafin}. The Plummer model requires spherical symmetry and the velocity distribution is assumed to be isotropic. In general, this is not likely to be true for dSphs \citep{1997NewA....2..139K}.
	\subsection{Newtonian ($p=2$) and Milgromian ($p=3$) limit for the velocity dispersion}
	In the Newtonian ($p=2$) and Milgromian ($p=3$) limit, Eq. \eqref{eq:sigmafin} reduces to the known expressions for the 1D velocity dispersion (cf. \citealt{2008gady.book.....B}) and \citealt{Safarzadeh_2021}):
	\begin{align}
		p=2 &\Rightarrow \sigma_\text{1D}^2 = \frac{\pi}{32} \frac{GM_\text{st}}{r_\text{pl}}\;,\\
		p=3 &\Rightarrow \sigma_\text{1D}^2 = \frac{2}{9} \sqrt{a_0GM_\text{st}}\;.
	\end{align}

	\section{Data analysis}
	\label{sec:ana}
	The data of Milky Way and Andromeda satellite dSphs from \citet{2017ApJ...836..152L} and a selection from the HIghest X-ray FLUx Galaxy Cluster Sample (HIFLUGS) from \citet{2023A&A...677A..24L} is used to check whether they hint at a relation between $p$ and the density of the satellite galaxy or the galaxy cluster. To accomplish this, what we interpret to be $g_\text{N}$ and $a$ from observations is obtained for each dSph and galaxy cluster. Since there is no interest in the radial dependence of $g_\text{N}(r)$ and $a(r)$, from now on $g_{\text{N}}$ and $a$ is written. With the RAR for arbitrary $p$ derived in Chapter \ref{sec:deriv}, $p$ is calculated. From these $p$-parameters, velocity dispersions are obtained for the Milky Way dSphs analysed. All relevant data are shown in Table \ref{tab:dsphs} for the dSphs and in Table \ref{tab:gcs} for the galaxy clusters. Figure \ref{fig:RARBoth} shows the data compared to the RAR for arbitrary $p$ derived in Chapter \ref{sec:deriv}. 
	
	For error propagation normally distributed and uncorrelated values are assumed. The error $\Delta \alpha$ of an arbitrary quantity $\alpha(\beta_1, ..., \beta_n)$ is calculated from arbitrary quantities $\beta_1, ..., \beta_n$, $n \in \mathbb{N}$ with uncertainties $\Delta\beta_1, ..., \Delta\beta_n$:
	\begin{equation}
		\label{eq:errorprop}
		\Delta \alpha = \sqrt{\sum_{i=1}^n \left(\frac{\partial \alpha}{\partial \beta_i}\Delta \beta_i\right)^2}\;.
	\end{equation}
	Uncertainties are calculated to two significant digits. If the third significant digit is smaller than 5, uncertainties are rounded down, otherwise rounded up.
	\subsection{RAR for the data at hand}
	What we interpret to be $g_{\text{N}}$ and $a$ of the dSphs has been inferred by \citet{2017ApJ...836..152L} from observational quantities. To accomplish this, they collected distances, V-band magnitudes, half-light-radii, and velocity dispersions from various sources. Velocity dispersions have been obtained from high-resolution individual-star spectroscopy. 
	Their values are adopted, interpreting their data the following way: 
	\begin{align}
		g_{\text{N}} &= g_{\text{bar}}\;,\\
		a &= g_{\text{obs}}\;.
	\end{align}
	They calculated $g_{\text{bar}}$ with Newton's law of gravity and stellar mass only contributing and $g_{\text{obs}}$ from velocity dispersion using an estimator by \citet{2010MNRAS.406.1220W}. Here, the analysis is limited to their high quality sample reducing the impact of tidal effects and neglecting galaxies with ellipticity larger than 0.45. For details see \citet{2017ApJ...836..152L}. 
	
	From the HIFLUGS clusters, the gas mass $M_{\text{gas},500}$, stellar mass $M_{*,500}$, and dynamical mass $M_{\text{dyn},500}$ within the virial radius $r_{500}$ are used. \citet{2023A&A...677A..24L} obtained the gas mass from integrating surface brightness fits and the gas mass by assuming a correlation to the stellar mass. They inferred the dynamical mass from X-ray and optical data, taking into account various morphological properties, dynamical properties and temperature, assuming hydrostatic equilibrium. They took their values for the virial radius from \citet{2011A&A...526A.105Z}, who obtained it from the X-ray measured mass distribution assuming hydrostatic equilibrium.
	
	The same derivation like \citet{2023A&A...677A..24L} is followed here. From the gas mass $M_{\text{gas},500}$, stellar mass $M_{*,500}$, and dynamical mass $M_{\text{dyn},500}$ enclosed in the virial radius $r_{500}$, for the galaxy clusters (gc) $a$ and $g_\text{N}$ are computed according to Newton's law of gravity:
	\begin{align}
		g_{\text{N,gc}} &= G \cdot \frac{M_{\text{gas},500}+{M_{*,500}}}{r_{500}^2}\;, \\
		a_{\text{gc}} &= G \cdot \frac{M_{\text{dyn},500}}{r_{500}^2}\;,
	\end{align}
	with the gravitational constant $G = \qty[exponent-product=\cdot]{4.3009172706(3)E-3}{\parsec\per\Msol(\kilo\metre \per\second)\squared}$. 
	Uncertainties were not given explicitly by \citet{2023A&A...677A..24L}.
	Uncertainties are taken for $M_{\text{gas},500}$ from \citet{2011A&A...526A.105Z} and calculated for $M_{*,500}$ according to Eq. (23) in \citet{2023A&A...677A..24L}:
	\begin{equation}
		\label{eq:stunc}
		\Delta M_{*,500} = \frac{4}{95} \cdot \left(\frac{\qty{5.7E13}{\Msol}}{M_{\text{gas},500}}\right)^{0.4} \cdot \Delta M_{\text{gas},500}\;.
	\end{equation}
	Uncertainties are not reconstructed for $r_{500}$ and $M_{\text{dyn},500}$, therewith underestimating the uncertainties for $g_{\text{N}}$ and $a$. The Uncertainty from the gravitational constant $G$ is neglected, since it is so small, that it would not contribute significantly to the uncertainties for $g_{\text{N}}$ and $a$.
	
\begin{figure}[ht!]
		\centering
		\includegraphics[width=0.5\textwidth]{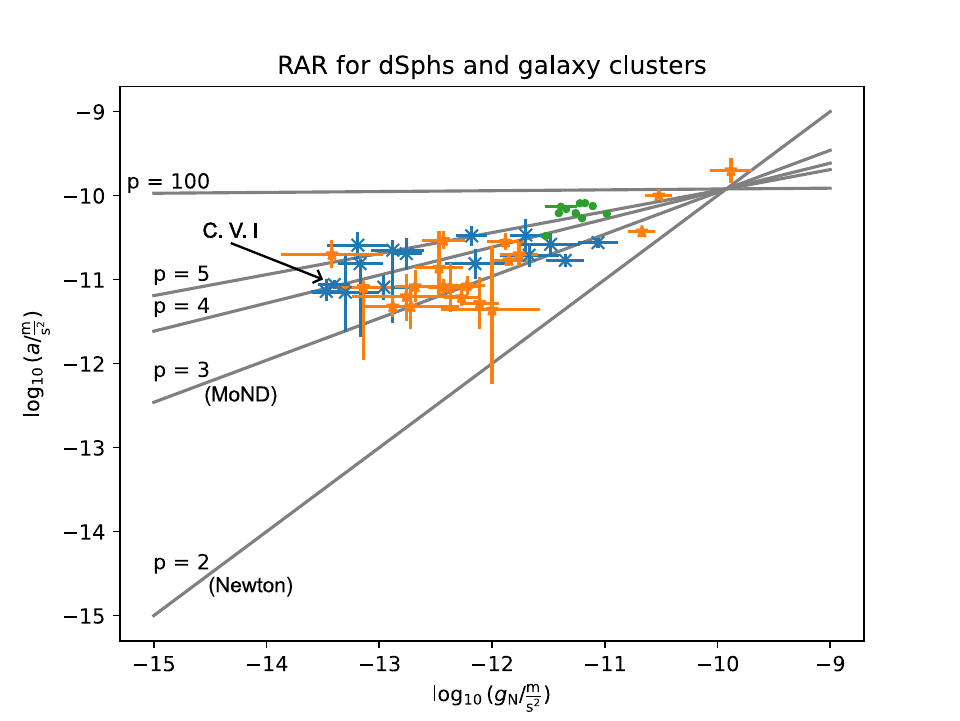}
		\caption{RAR for selected Milky Way (blue) and Andromeda (orange) satellite dSphs and galaxy clusters (green). Uncertainties are $1\sigma$. The data are taken from \citet{2017ApJ...836..152L} for the dSphs (Milky Way and Andromeda satellites) and \citet{2023A&A...677A..24L} for the galaxy clusters, as tabulated in Tables \ref{tab:dsphs} and \ref{tab:gcs}, respectively. The bold grey lines for different $p$ are the same as in Fig. \ref{fig:1}.}
		\label{fig:RARBoth}
	\end{figure}

	Figure \ref{fig:RARBoth} shows the RAR for the data at hand. Deviations from the $p=3$ line can clearly be seen. For Canes Venatici I the deviation reaches a $5\sigma$-significance. The galaxy cluster data show a $5\sigma$-deviation at least from the $p=3$ limit subject to the caveat that their uncertainties are underestimated.
	
	\subsection{Calculating $p$ from the RAR and density for the data at hand}
	To calculate $p$ for the data at hand Eq. \eqref{eq:prar} is used. Solving Eq. \eqref{eq:prar} for $p$ yields:
	\begin{equation}
		p = \frac{\log_{10}\left(\frac{g_\text{N}}{a_0}\right)}{\log_{10}\left(\frac{a}{a_0}\right)} + 1 \label{eq:p}\;.
	\end{equation}
	Error propagation according to Eq. \eqref{eq:errorprop} demands symmetrical uncertainties, which is not fulfilled the values for $a$ and $g_\text{N}$. Uncertainties are therefore symmetrised by taking the larger uncertainty, overestimating the uncertainties. The error of $a_0$ does not significantly influence the uncertainties of $p$ for the dSphs, but for galaxy clusters. This is because not all relevant uncertainties for the galaxy clusters are considered.
	The baryonic mass of the dSphs is computed from their V-band luminosity $L_{\text{V}}$ and stellar mass-to-light-ratio $\Upsilon_* = \qty{2.0(5)}{\Msol\per\Lsol}$ given by \citet{2017ApJ...836..152L}:
	\begin{equation}
		M_{\text{dSphs}} = \Upsilon_* \cdot L_{\text{V}}\;.
	\end{equation}
	\citet{2017ApJ...836..152L} inferred the mass-to-light-ratio as a statistical mean from studies of Fornax and Sculptor dSphs by \citet{2012A&A...544A..73D}.
	The mean density $\bar{\rho}_{\text{dSphs}}$ of the dSphs is estimated from their half-light-radius $r_{1/2}$ from \citet{2017ApJ...836..152L}:
	\begin{equation}
		\bar{\rho}_{\text{dSphs}} = \frac{4\pi}{3} \cdot \frac{M_{\text{dSphs}}}{2r_{1/2}^3}\;.
	\end{equation}
	The mean density $\bar{\rho}_{\text{gc}}$ of the galaxy clusters is estimated as:
	\begin{equation}
		\bar{\rho}_{\text{gc}} = \frac{4\pi}{3} \cdot \frac{M_{\text{gas},500}+M_{*,500}}{r_{500}^3}\;.
	\end{equation}
	The exact shape is neither considered for the dSphs nor the galaxy clusters which leads to systematic uncertainty. 
	
\begin{figure}[ht!]
		\centering
		\includegraphics[width=0.5\textwidth]{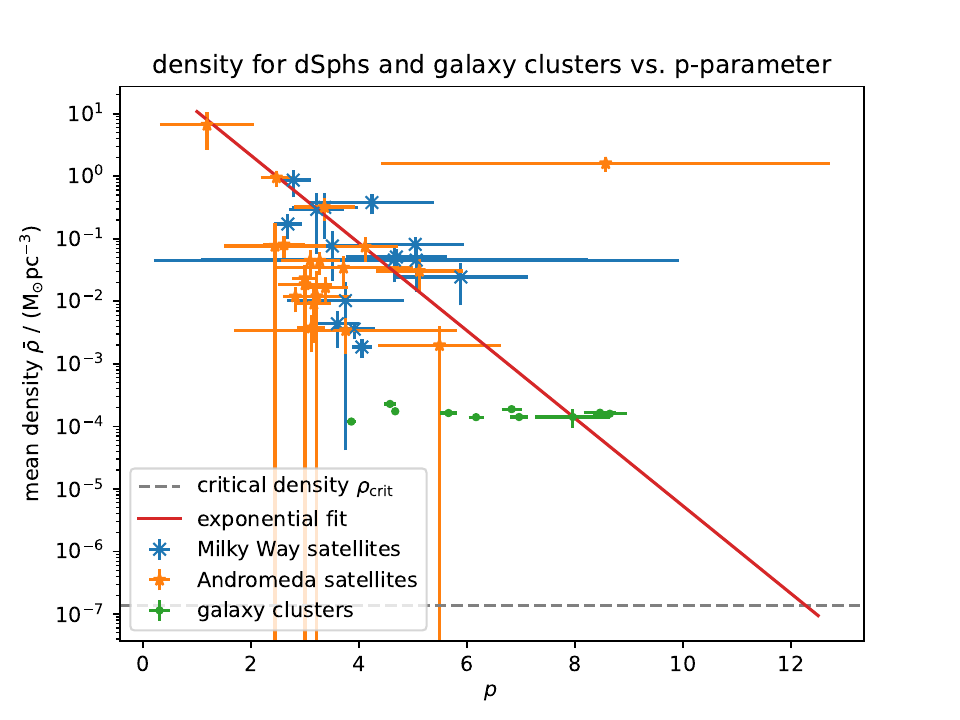}
		\caption{Relation between density of dSphs (blue and orange) and galaxy clusters (green) and $p$-parameter from GPE, Eq. \eqref{eq:plapl}. Uncertainties are $1\sigma$. $p$ is obtained from the RAR as shown in Fig. \ref{fig:RARBoth}. See Tables \ref{tab:dsphs} and \ref{tab:gcs} for values. The red line is an exponential function fit to the data with $\chi^2/\text{ndf}=8.001$.}
		\label{fig:pdensity}
	\end{figure}

	Figure \ref{fig:pdensity} shows the density of the analysed dSphs and galaxy clusters against $p$. The $p$-parameters deviate from $p=3$, in many cases with $1\sigma$-significance, for some with $2\sigma$-significance, for Canes Venatici I with $5\sigma$-significance. The uncertainties on $L_\text{V}$ and $r_{1/2}$ and therefore on $p$ of Canes Venatici I are relatively small. For the galaxy clusters, we refrain from making a claim, since the uncertainties are underestimated. Furthermore, the python package kafe2 is used to fit an exponential function of the following form and with the parameters $\rho_{\alpha}$ and $\alpha$ to the data with ndf being the number of degrees of freedom:
	\begin{align}
		\rho(p) &= \rho_{\alpha} \cdot \exp(\alpha \cdot p)\; \label{eq:rhop}, \\
		\rho_{\alpha} &= \qty{54(19)}{\Msol\parsec^{-3}}\;, \\
		\alpha &= \num{-1.614(43)}\;,\\
		\chi^2/\text{ndf} &= 8.001\;,\\
		\text{ndf} &= 43\;. 
	\end{align} 
	$\qty{8}{\Msol\parsec^{-3}}$ is used as a starting value for $\rho_{\alpha}$. The large $\chi^2/\text{ndf}$ is due to the small uncertainties for the galaxy clusters which are underestimated. Without the galaxy clusters, $\chi^2/\text{ndf}=3.115$ is obtained.\footnote{With the same starting value, but different fit parameters: $\rho_{\alpha} = \qty{3.2(68)}{\Msol\parsec^{-3}}$ and $\alpha = \qty{-1.81(60)}{}$, $\text{ndf} = 33$.} This behaviour is extrapolated to the critical density $\rho_{\text{crit}}$ of the universe at redshift $z=0$ from the WMAP 7-year cosmology implemented in the astropy.cosmology python package:
	
	\begin{align}
		\rho_{\text{crit}} &\approx \qty{1.38e-7}{\Msol\parsec^{-3}}\;,\\
		p(\rho_{\text{crit}}) &= \num{12.27(39)}\;.
	\end{align}
	
	\subsection{Infinity Laplacian for weak fields}
	When $p \rightarrow{} \infty$, the $p$-Laplace operator transitions to a limiting form. The non-linear operator, $-\nabla \cdot \left( |\nabla \phi|^{p-2} \nabla \phi \right)$, converges to the infinity Laplacian \citep{lindqvist2015notesinfinitylaplaceequation}:
	\begin{equation}
		\Delta_\infty \phi = \nabla \phi \cdot \nabla \left( \nabla \phi \right) =\sum_{i,j} \frac{\partial \phi}{\partial x_i} \frac{\partial^2 \phi}{\partial x_i \partial x_j} \frac{\partial \phi}{\partial x_j},
	\end{equation}
	\noindent where $\nabla (\nabla \phi)$ represents the Hessian matrix of second derivatives. The indices  $i$  and  $j$  range over the spatial dimensions ($n=3$ in our case), and $\partial$ denotes partial derivatives with respect to the spatial coordinates. The infinity Laplacian governs the minimization of the maximum gradient of the potential $\phi$, known as the absolutely minimal Lipschitz extension problem \citep{CAMILLI201771}. In practice, when $p \rightarrow{} \infty$, the infinity-Laplace operator smooths out the steepest gradients \citep{Lindqvist2019}, resulting in an acceleration of nearly constant magnitude, consistent with Eq. (9). This result independently suggests that in the weakest gravitational potentials, including well beyond $\rho_{\text{crit}}$, the GPE predicts that gravity would smooth out any sharp variations in magnitude that deviate from  $a_0$,  across the whole non-linear domain.

	\subsection{Limitations of this analysis} 
	The present work serves as an invitation for future observational studies, which are within reach instrumentally but have yet to undergo systematic and thorough investigation. To reduce the uncertainty for the mass-to-light-ratios of the dSphs, we suggest an analysis which considers individual mass-to-light-ratios for the dSphs. This could be achieved within a homogeneous study by one single team. In such an analysis, uncertainties on $a$ and $g_\text{N}$ might also be reduced. To estimate the density more accurately, we suggest considering the shape of the individual objects analysed. That the half-light-radius is used for the dSphs and the virial radius for the galaxy clusters, might complicate comparability. We also suggest in the future an analysis with all uncertainties for the galaxy clusters. These improvements were here not possible as the corresponding data are not available. 
	
	Finally, the present work constitutes a static analysis of the dynamics generated by the GPE, Eq. \eqref{eq:plapl}. At present, live simulations of the studied objects are only possible for the $p=2$ and $p=3$ cases (e.g. using the code Phantom of Ramses, \citet{2015CaJPh..93..232L}). Simulations of the satellite galaxies orbiting their host galaxy can thus be performed for these values of $p$. For example, a detailed study for $p=2$ by \citet{1997NewA....2..139K} already suggested that even Newtonian solutions without dark matter exist for most of the galaxy's satellites. It will be important to extend such work to the $p=3$ case to study if apparent $p>3$ solutions emerge.
	
	\subsection{Calculating the velocity dispersion for the Plummer model from an obtained $p$}
\begin{figure}[ht!]
		\centering
		\includegraphics[width=0.5\textwidth]{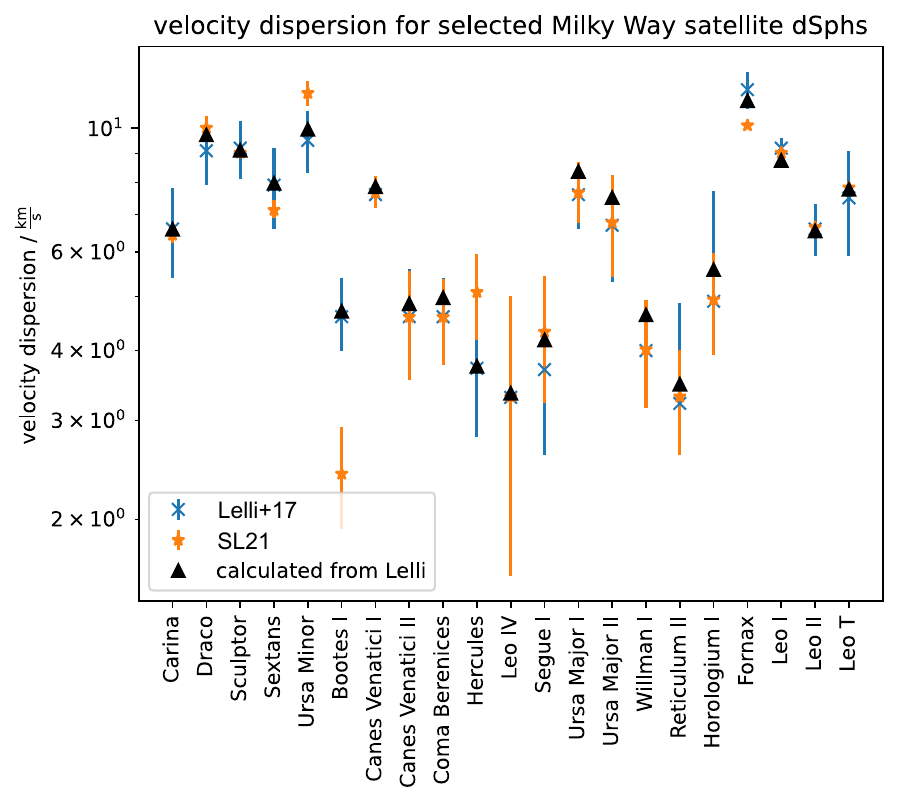}
		\caption{1D velocity dispersion for selected Milky Way satellite dSphs. Blue are data by \citet{2017ApJ...836..152L} and orange data by \citet{Safarzadeh_2021}. Black triangles are calculated from the \citet{2017ApJ...836..152L} data with Eq. \eqref{eq:sigmafin}.}
		\label{fig:sigma}
	\end{figure}

	To infer velocity dispersions of the dSphs from the data by \citet{2017ApJ...836..152L}, Eq. \eqref{eq:sigmafin} is used, identifying $r_\text{pl} = \frac{r_{1/2}}{1.305}$ \citep{Kroupa_2008} and $M_\text{st} = M_\text{dSphs}$. Now, the full sample of Milky Way satellites is used and compared to the \citet{2017ApJ...836..152L} and \citet{Safarzadeh_2021} data neglecting errors in Fig. \ref{fig:sigma}. The sample of Milky Way satellites chosen by \citet{Safarzadeh_2021} is used. The latter adopted velocity dispersion data from various astrophysical literature sources. Any calculated velocity dispersion lies within the $1\sigma$-regime of the corresponding data point by \citet{2017ApJ...836..152L} and often also of the corresponding \citet{Safarzadeh_2021} point. The outliers are Ursa Minor with a 2$\sigma$-tension and Bootes I as well as Fornax with 3$\sigma$-tension between the \citet{2017ApJ...836..152L} and \citet{Safarzadeh_2021} points. The ellipticity of Ursa Minor is 0.54, which might be not negligible. Still, for Ursa Major I with an ellipticity of 0.80, there is no tension at all. Bootes I has no peculiar properties, there might be systematic uncertainty not considered. Fornax is the most massive Milky Way dSph from the sample at hand.
	\section{Interpreting GPE as modification of gravitational acceleration under the influence of external Newtonian gravity}
	\label{sec:ext}
\begin{figure}[ht!]
		\centering
		\includegraphics[width=0.5\textwidth]{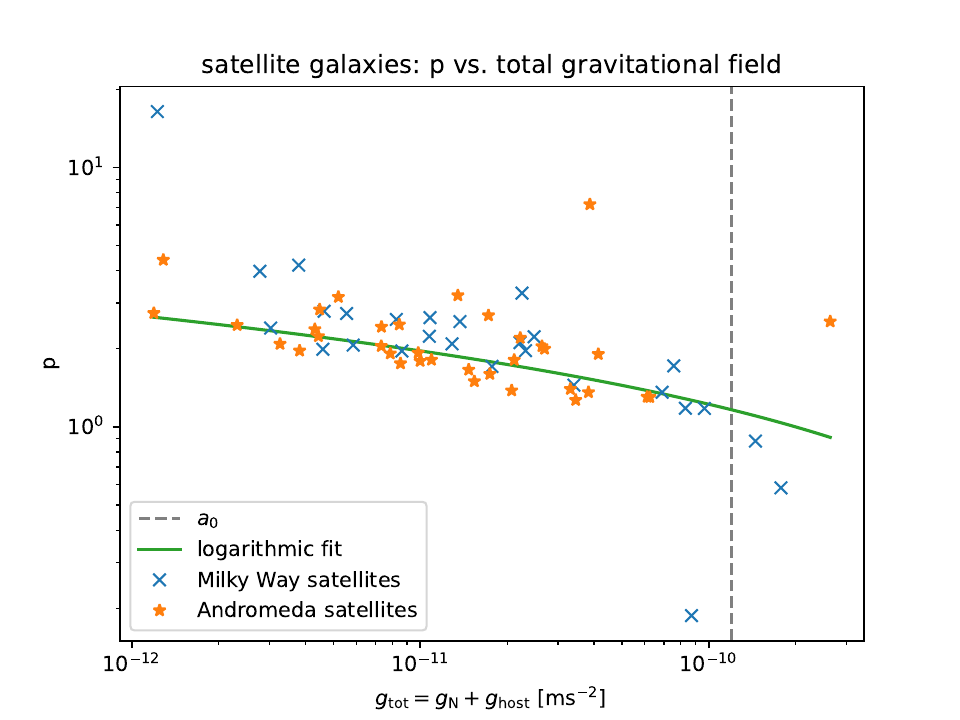}
		\caption{Relation between $p$-parameter and external field $g_\text{tot}$ for Milky Way (blue) and Andromeda (orange) satellite dSphs. The dashed line describes Milgrom's constant $a_0$. Uncertainties are not shown to keep the plot legible (see Tables \ref{tab:mwext} and \ref{tab:andext} for uncertainties on $p$). The external field is Newtonian and consists of the field by the host galaxy and the internal field of the satellite. A logarithmic fit function (green solid curve) described by Eq. \eqref{eq:fitmodel} suggests a functional relation $p(g_\text{tot})$.}
		\label{fig:accgtot}
	\end{figure}

	In this section, the following hypothesis is examined: The GPE and MoND are not a modification of fundamental dynamics, but just a modification of acceleration. Drawing an analogy to electrorheological fluids, suggests gravity remains Newtonian on a fundamental level, but the $p$-parameter describes a non-linear response of the 'induced' field $\vec{a}(r)$ inside the satellite galaxy to an external Newtonian field $\vec{g}_\text{host}(r)$ by the host galaxy and $\vec{g}_\text{N}(r)$ by the satellite itself. This would mean, that the Poisson Equation still holds true while the gravitational field is modified:
	\begin{align}
		\nabla \cdot \vec{g}_\text{tot} = 4\pi G \rho_\text{tot}(r)\;, \\
		\vec{g}_\text{tot} := \vec{g}_\text{host}(r) + \vec{g}_\text{N}(r) = \left(\frac{|\vec{a}(r)|}{a_0}\right)^{p-2} \vec{a}(r) \label{eq:gtot} \;.
	\end{align}
	In this picture, $\vec{g}_\text{tot}$ takes the role of electric field strength $\vec{E}$ and $\vec{a}$ the role of electric displacement $\vec{D}$. The $p$-parameter is used to describe viscosity behavior of electrorheological fluids. An analogue physical process in galaxies is unknown. The theory of electrorheological fluids postulates a functional relation between the $p$-parameter and the external field. Here we likewise inquire whether an analogue relation might exist for the analyzed dSphs. For this task, the $p$-parameter is reevaluated:
	\begin{align}
		p &= \frac{\log_{10}\left(\frac{g_\text{tot}}{a_0}\right)}{\log_{10}\left(\frac{a}{a_0}\right)} + 1\;, \label{eq:pgtot} \\
		g_\text{tot} &:= |\vec{g_\text{tot}}| = |\vec{g}_\text{N} + \vec{g}_\text{host}| = g_\text{N} + g_\text{host} \;, \label{eq:gtotadd}\\
		g_\text{host} &:= \frac{GM_\text{host}}{D_\text{host}^2}. \label{eq:ghost}
	\end{align} 
	Equation \eqref{eq:gtotadd} holds thanks to superposition and spherical symmetry. $M_\text{host}$ is the mass of the host galaxy ($M_\text{host} = \qty{1E12}{\Msol}$ for the Milky Way and $M_\text{host} = \qty{2E12}{\Msol}$ for Andromeda). $D_\text{host}$ is the distance of the satellite to its host. All values are again taken from \citet{2017ApJ...836..152L}. Since no uncertainties for $M_\text{host}$ and $D_\text{host}$ were given, the uncertainties for $g_\text{tot}$ are the same as for $g_\text{N}$. The uncertainty on $G$ is still neglected. For this task all 60 satellites analyzed by \citet{2017ApJ...836..152L} are investigated.
	A logarithmic fit leads to the following result in Fig. \ref{fig:accgtot}:
	\begin{align}
		p &= p_0 \cdot \log_{10}\Bigg(\frac{g_\text{tot}}{g_0}\Bigg) + 1 \label{eq:fitmodel} \;, \\
		p_0 &= \num{-0.74(43)}\;,\\
		g_0 &= \num{2.0(40)E-10}\unit{\metre\per\second\squared}\;,\\
		\chi^2/\text{ndf} &= 0.02892\;, \\
		\text{ndf} &= 60\;.
	\end{align}
	The fit model (Eq. \eqref{eq:fitmodel}) is chosen to be identical to Eq. \eqref{eq:pgtot}, if $a_0=g_0$ and $p_0 = \frac{1}{\log_{10}\left(\frac{a}{a_0}\right)}$. Due to large uncertainties, these identities cannot be confidently defended. This would imply $a \propto a_0$. Extrapolating to large $g_\text{tot}$ is unphysical, since Eq. \eqref{eq:fitmodel} does not reproduce Newtonian behavior and allows $p \leq 1$, which is mathematically forbidden \citep{LindqvistpLaplace}. Still, extrapolating to low accelerations yields a testable prediction. The small $\chi^2/\text{ndf}$ is due to large uncertainties. There seem to be five outliers including two Milky Way dSphs with $p \leq 1$. Still, uncertainties are so large, that they cannot be identified as such.

	\section{Discussion}
	\label{sec:dis}
	This analysis leads to the following conclusions:
	\begin{enumerate}
		\item The gravitational acceleration $a(r)$ generated by the source distribution can be given by Eq. \eqref{eq:acc}, derived from the GPE, Eq. \eqref{eq:plapl}, with $p$ depending on $\bar{\rho}$ according to Eq. \eqref{eq:rhop}. \footnote{Item \ref{item:1} implies the present model comprises energy conservation. Since the model is assumed to be static, no claims about conservation of momentum can be stated.}\label{item:1} 
		\item For large $p$ the Milgromian gravitational acceleration $a$ becomes independent of central mass and distance and equals $a_0$ (Eq. \ref{eq:ainf}). \label{item:2}
		\item The analysed data hint at a relation of $p$ and the density of the bound system $\bar{\rho}$ (Eq. \eqref{eq:rhop}). The lower the density, the higher $p$. For the critical density  $p(\rho_{\text{crit}}) = \num{12.27(39)}$ is found. \label{item:3}
		\item Combining the GPE with a Plummer model predicts the observed velocity dispersions in Milky Way dSphs with less than 1$\sigma$ deviation without exception (Fig. \ref{fig:sigma}). That is, once the density of the satellite galaxy is known we can determine the $p$-parameter using Eq. \eqref{eq:rhop} and from Eq. \eqref{eq:sigmafin} we obtain an estimate for the velocity dispersion. \label{item:4}
		\item Interpreting the GPE as fundamentally Newtonian, but the gravitational field $a$ as a non-linear response to an external field $g_\text{tot}$ requires a relation $p(g_\text{tot})$. The analysed data hints at such a relation (Fig. \ref{fig:accgtot}). \label{item:5}
	\end{enumerate}
	Items \ref{item:2} and \ref{item:3} imply that for densities of the bound system on cosmological scales the Milgromian gravitational acceleration $a$ becomes non-directional and independent of central mass and distance and equals $a_0$. This increases the gravitational acceleration compared to Newtonian gravity reducing the need for dark matter. Free particles are not possible for $p>3$, since $\lim\limits_{r \to \infty} \Phi(r)$ does not exist. The exponential fit (Eq. \eqref{eq:rhop}) cannot be extrapolated to densities related to $p<2$, because it is known there is Newtonian gravitation ($p=2$) for far larger densities than $\rho(p=2) = \qty{2.14(78)}{\Msol\parsec^{-3}}$.
	
	Items \ref{item:1} and \ref{item:2} are derived from the model presented in Sect. \ref{sec:deriv}. This model makes many simplifying assumptions (e.g. spherical symmetry) discussed in Sect. \ref{sec:deriv}. Item \ref{item:3} is suggested by the data presented in Sect. \ref{sec:ana}. From the data claims of up to $5\sigma$-significance are stated. 
	
	Item \ref{item:4} might solve the problem of MoND not being able to predict all velocity dispersions in Milky Way satellites raised by \citet{Safarzadeh_2021}. It must be considered, that the $p$-parameters have been inferred from the observed accelerations, which have been inferred from velocity dispersions by \citep{2017ApJ...836..152L}. An independent test might be beneficial to check whether the data correctly represent the Milgromian acceleration $a$, which is fundamentally independent of velocity dispersion.
	
	Apart from items \ref{item:3} and \ref{item:5}, the physical meaning of $p$ is to be explored. $p$ might be explained by drawing the parallel to electrorheological fluids. This has been done for standard MoND by \citet{Blanchet_2007}. The data suggests a relation $p(g_\text{tot})$, which makes a prediction for low accelerations. Still, an analysis with more precise data is desirable. If the analogy to electrorheological fluids was justified, any data analysis would always have to explore the effect of external fields. Furthermore, this raises the question whether there is a physical process leading to non-linear response of dSphs to external gravitational fields. The analysis presented here may suggest that the GPE describes an effective coupling between the dynamics of a low-acceleration system to the gravitational field in the non-relativistic limit.
	
	This analysis is yet one more hint that there might be a deeper underlying gravitational theory generalizing MoNDian and Newtonian gravity. This justifies to further review current cosmological models and gravitational theories.

	\bibliographystyle{bibtex/aa}
	\bibliography{bibtex/literature}
	\begin{appendix}
		\onecolumn
		\section{RAR data}
		\label{sec:app}
		\begin{table*}[h!] 
			\centering
			\caption{All relevant quantities for the dSphs and galaxy clusters.}
			\label{tab:dsphs}
			\begin{tabular}{l|S[table-format=1.2(2)]|S[table-format=4.0(3)]|S[table-format=-2.2(2)]|S[table-format=-2.2(2)]|S[table-format=1.2(2)]|S[table-format=3.3(6)]}
				Galaxy & {$\log_{10}(\frac{L_{\text{V}}}{\text{L}_{\odot}})$} & {$\frac{r_{1/2}}{\unit{\parsec}}$} & {$\log_{10}(\frac{g_{\text{N}}}{\unit{\metre\per\second\squared}})$} & {$\log_{10}(\frac{a}{\unit{\metre\per\second\squared}})$} & {$p$} & {$\frac{\bar{\rho}}{10^{-2}\unit{\Msol\parsec^{-3}}}$} \\
				\midrule
				Carina & 5.57(20) & 273(45) & -12.15(27) & -10.81(18) & 3.51(59) & 7.6(55) \\ 
				Draco & 5.45(8) & 244(9) & -12.18(14) & -10.48(12) & 5.04(90) & 8.1(27)  \\ 
				Fornax & 7.31(12) & 792(58) & -11.35(17) & -10.77(8) & 2.68(26) & 17.2(7.4)  \\ 
				Leo I & 6.74(12) & 298(29) & -11.06(18) & -10.56(6) & 2.78(33) & 87(41)  \\ 
				Leo II & 5.87(12) & 219(52) & -11.67(26) & -10.71(14) & 3.22(51) & 30(24)  \\ 
				Sculptor & 6.36(20) & 311(46) & -11.48(26) & -10.58(13) & 3.37(61) & 32(22)\\
				Sextans & 5.64(20) & 748(66) & -12.96(24) & -11.09(15) & 3.59(39) & 0.44(0.26) \\
				Bootes I & 4.29(8) & 283(7) & -13.47(14) & -11.14(15)\tablefootmark{*} & 3.91(38) & 0.36(0.12) \\
				Canes V. I & 5.08(8) & 647(27) & -13.40(14) & -11.06(5) & 4.05(18) & 0.186(0.062) \\
				Canes V. II & 4.10(8) & 101(5) & -12.76(14) & -10.69(19) & 4.69(93) & 5.1(1.8) \\
				Coma B. & 3.46(24) & 79(6) & -13.19(27) & -10.59(16) & 5.9(12) &  2.5(1.6) \\
				Hydra II & 3.87(12) & 88(17) & -12.88(28) \tablefootmark{*} & -10.65(87) \tablefootmark{*} & 5.1(49) & 4.6(3.1) \\
				Leo IV & 3.91(8) & 149(47) & -13.30(30) & -11.15(47) & 3.8(11) &  1.0(1.0) \\
				Leo T & 5.57(7) & 160(10) & -11.70(14) & -10.47(19) & 4.2(12) & 38(13)  \\
				Segue II & 2.94(12) & 43(6) & -13.17(20) & -10.81(87) & 4.7(36) & 4.6(2.6) \\
				\midrule
				NGC 147 & 7.84(4) & 672(23) & -10.67(12) & -10.43(6) & 2.47(29) & 96(27)  \\
				NGC 185 & 7.84(4) & 565(7) & -10.52(12) & -10.00(4) & 8.6(42) & 161(43) \\
				NGC 205 & 8.52(4) & 594(107) & -9.88(19) & -9.70(15) & 1.18(87) & 660(400)  \\
				And II & 6.85(8) & 1355(142) & -12.27(16) & -11.21(11) & 2.82(20) & 1.19(0.53)  \\
				And V & 5.55(8) & 365(57) & -12.43(19) & -10.53(11) & 5.12(81) & 3.1(1.7)  \\
				And VI & 6.44(8) & 537(54) & -11.88(16) & -10.55(11) \tablefootmark{*}& 4.11(60) & 7.5(3.2)  \\
				And VII & 6.98(12) & 965(52) & -11.85(17) & -10.77(7) & 3.27(27) & 4.5(1.8)  \\
				And IX & 4.97(44) & 582(23) & -13.42(45) & -10.70(17)\tablefootmark{*} & 5.5(11) & 0.20(0.21)  \\
				And X & 5.13(40) & 312(41) & -12.72(43) & -11.32(27) & 3.00(49) & 1.9(1.9)  \\ 
				And XIV & 5.37(20) & 434(212) & -12.76(48) & -11.20(29) \tablefootmark{*}& 3.22(63) & 1.2(1.9)  \\ 
				And XV & 5.69(12) & 294(12) & -12.11(17) & -11.28(31) & 2.61(39) &  8.1(3.2)\\ 
				And XVI & 5.53(12) & 164(9) & -11.76(17) & -10.70(17) \tablefootmark{*}& 3.36(56) & 32(13)  \\ 
				And XVII & 5.34(16) & 299(31) & -12.47(21) & -10.86(44) \tablefootmark{*} & 3.7(13) & 3.4(1.9)  \\ 
				And XXIII & 6.00(20) & 1034(97) & -12.88(24) & -11.32(13) & 3.11(26) & 0.38(0.23)  \\ 
				And XXIV & 5.32(20) & 633(52) & -13.14(24) & -11.09(87) \tablefootmark{*}& 3.8(21) & 0.35(0.20)  \\ 
				And XXVIII & 5.39(16) & 285(12) & -12.37(20) & -11.09(29) & 3.09(55) & 4.4(2.1)  \\ 
				And XXIX & 5.33(12) & 377(30) & -12.68(18) & -11.08(19) & 3.38(42) & 1.67(0.74)  \\
				And XXXI & 6.61(28) & 1231(132) & -12.43(32)\tablefootmark{*} & -11.08(9) & 3.16(32) & 0.91(0.70)  \\
				Cetus & 6.44(8) & 791(75) & -12.22(16) & -11.07(11) & 3.00(24) & 2.33(0.98)  \\
				Perseus I & 6.04(28) & 391(143) & -12.00(45)\tablefootmark{*} & -11.36(88) \tablefootmark{*} & 2.44(94) & 8(10)  \\
			\end{tabular}
			\tablefoot{Above the horizontal line are Milky Way satellites, below it are Andromeda satellites. The columns are from left to right: name of the galaxy, V-band-luminosity $L_\text{V}$, half-light-radius $r_{1/2}$, Newtonian acceleration $g_\text{N}$, observationally deduced acceleration $a$, p-parameter from p-Laplacian $p$ (Eq. \eqref{eq:plapl}), mean density $\bar{\rho}$. The first four quantities are adopted from \citet{2017ApJ...836..152L} and $p$ and $\bar{\rho}$ are calculated from them. \tablefoottext{*}{Uncertainty deviates from data by \citet{2017ApJ...836..152L}, because it has been symmetrised.}}
	
			\label{tab:gcs}
			\begin{tabular}{
					l|
					S[table-format=4.0]|
					S[table-format=2.2(2)]|
					S[table-format=1.2(2)]|
					S[table-format=2.2]|
					S[table-format=-2.3(4)]|
					S[table-format=-2.4]|
					S[table-format=1.3(2)]|
					S[table-format=1.3(2)]
				}
				{Cluster} & {$\frac{r_{500}}{\si{kpc}}$} & {$\frac{M_{\text{gas},500}}{10^{13}\si{\Msol}}$} & {$\frac{M_{*,500}}{10^{12}\si{\Msol}}$}& {$\frac{M_{\text{dyn},500}}{10^{14}\si{\Msol}}$}& {$\log_{10}\left(\frac{g_{\text{N}}}{\si{\metre\per\second\squared}}\right)$} & {$\log_{10}\left(\frac{a}{\si{\metre\per\second\squared}}\right)$} & {$p$} & {$\frac{\bar{\rho}}{10^{-4}\si{\Msol\parsec^{-3}}}$} \\
				\midrule
				A0085 & 1217 & 6.67(32) & 4.39(13) & 8.66 & -11.175(20) & -10.089 & 8.46(30) & 1.652(74) \\
				A0262 & 755 & 1.08(12) & 1.47(10) & 1.35 & -11.523(43) & -10.481 & 3.858(80) & 1.19(12) \\
				A0496 & 967 & 2.79(12) & 2.60(7) & 4.65 & -11.342(17) & -10.159 & 6.97(17) & 1.413(56) \\
				A0576 & 869 & 2.00(71) & 2.13(45) & 4.00 & -11.39(14) & -10.132 & 7.96(69) & 1.413(45) \\
				A1795 & 1085 & 4.95(14) & 3.68(6) & 4.54 & -11.201(11) & -10.270 & 4.670(64) & 1.744(46) \\
				A2029 & 1247 & 8.24(29) & 4.99(11) & 8.38 & -11.106(14) & -10.124 & 6.83(19) & 1.887(63) \\
				A2142 & 1371 & 13.40(80) & 6.68(24) & 8.17 & -10.982(25) & -10.218 & 4.57(10) & 2.29(13) \\
				A2589 & 837 & 1.77(12) & 1.98(8) & 3.11 & -11.407(27) & -10.208 & 6.17(13) & 1.406(86) \\
				A3158 & 1013 & 3.75(33) & 3.11(16) & 4.56 & -11.258(35) & -10.208 & 5.66(15) & 1.64(13) \\
				A3571 & 1133 & 5.16(28) & 3.77(12) & 7.47 & -11.221(22) & -10.091 & 8.64(31) & 1.595(81) \\
			\end{tabular}
			\tablefoot{The columns are from left to right: name of the cluster, virial radius $r_{500}$, gas mass $M_{\text{gas},500}$, stellar mass $M_{*,500}$, dynamical mass $M_{\text{dyn},500}$, Newtonian acceleration $g_{\text{N}}$, observationally deduced acceleration $a$, p-parameter from p-Laplacian $p$ (Eq. \eqref{eq:plapl}), mean density $\bar{\rho}$. We adopted the first four quantities from \citet{2023A&A...677A..24L} and calculated from them $g_{\text{N}}$, $a$,  $p$ and $\rho$. Uncertainties for $M_{\text{gas},500}$ were taken from \citet{2011A&A...526A.105Z}. Uncertainties for $M_{*,500}$ have been reconstructed via Eq. \eqref{eq:stunc}.}
		\end{table*}
		\FloatBarrier
		\twocolumn
		
		\section{Velocity Dispersion}
		\begin{table}[h!]
			\caption{Velocity dispersion for Milky Way dSphs.}

			\sisetup{
				table-alignment-mode = marker
			} 
			\begin{tabular}{l|r|r|r}
				Galaxy &{Lelli+17 (1)}&{SL21 (2)}&{this work (3)}\\
				&{$[\unit{\kilo\metre\per\second}]$}&{$[\unit{\kilo\metre\per\second}]$}&{$[\unit{\kilo\metre\per\second}]$} \\
				\midrule
				Carina & $6.6^{+1.2}_{-1.2}$ & $6.44^{+0.20}_{-0.26}$ & \num{6.59} \\
				Draco & $9.1^{+1.2}_{-1.2}$ & $10.0^{+0.53}_{-0.50}$ & \num{9.72} \\
				Sculptor & $9.2^{+1.1}_{-1.1}$ & $9.03^{+0.19}_{-0.18}$ & \num{9.12} \\
				Sextans & $7.9^{+1.3}_{-1.3}$ & $7.13^{+0.22}_{-0.29}$ & \num{7.96} \\
				Ursa Minor & $9.5^{+1.2}_{-1.2}$ & $11.54^{+0.61}_{-0.58}$ & \num{9.94} \\
				Bootes I & $4.60^{+0.80}_{-0.60}$ & $2.41^{+0.49}_{-0.51}$ & \num{4.70} \\
				Canes V. I & $7.60^{+0.40}_{-0.40}$ & $7.66^{+0.49}_{-0.53}$ & \num{7.85} \\
				Canes V. II & $4.6^{+1.0}_{-1.0}$ & $4.59^{+1.04}_{-0.96}$ & \num{4.85} \\
				Coma B. & $4.60^{+0.80}_{-0.80}$ & $4.59^{+0.82}_{-0.77}$ & \num{4.97} \\
				Hercules & $3.70^{+0.90}_{-0.90}$ & $5.09^{+0.91}_{-0.86}$ & \num{3.75} \\
				Leo IV & $3.3^{+1.7}_{-1.7}$ & $3.3^{+1.7}_{-1.7}$ & \num{3.36} \\
				Segue I & $3.7^{+1.4}_{-1.1}$ & $4.3^{+1.1}_{-1.1}$ & \num{4.18} \\
				Ursa Major I & $7.6^{+1.0}_{-1.0}$ & $7.66^{+0.91}_{-1.02}$ & \num{8.36} \\
				Ursa Major II & $6.7^{+1.4}_{-1.4}$ & $6.8^{+1.4}_{-1.5}$ & \num{7.51} \\
				Willman I & $4.00^{+0.80}_{-0.80}$ & $4.02^{+0.86}_{-0.91}$ & \num{4.63} \\
				Reticulum II & $3.2^{+1.6}_{-0.5}$ & $3.31^{+0.71}_{-0.69}$ & \num{3.48} \\
				Horologium I & $4.9^{+2.8}_{-0.9}$ & $4.9^{+1.0}_{-1.0}$ & \num{5.58} \\
				Fornax & $11.70^{+0.90}_{-0.90}$ & $6.99^{+0.91}_{-1.06}$ & \num{11.18} \\
				Leo I & $9.20^{+0.40}_{-0.40}$ & $8.7^{+2.6}_{-2.7}$ & \num{8.75} \\
				Leo II & $6.60^{+0.70}_{-0.70}$ & $4.3^{+0.98}_{-1.01}$ & \num{6.54} \\
				Leo T & $7.50^{+1.60}_{-1.60}$ & $10.1^{+0.00}_{-0.00}$ & \num{7.77} \\
			\end{tabular}
			\tablebib{(1)~\citet{2017ApJ...836..152L}; (2) \citet{Safarzadeh_2021}; (3) calculated from data by \citet{2017ApJ...836..152L} with Eq. \eqref{eq:sigmafin}}
		\end{table}

		\begin{table}[h!]
			\section{external field Milky Way}
			\caption{Data on how Milky Way affects its dSphs via gravity.}
			\label{tab:mwext}
			\begin{tabular}{l|r|r|r|r}
				Galaxy &{$\frac{D_\text{host}}{ \unit{\kilo\parsec}}$}&{$\log_{10}(\frac{g_\text{host}}{\unit{\metre\per\second\squared}})$}& {$\log_{10}(\frac{g_\text{N}}{\unit{\metre\per\second\squared}})$} & {$p$}\\
				\midrule
				Carina & $107$ & $-10.91$ & $-12.15 \pm 0.27$ & $2.1 \pm 2.3$ \\
Draco & $76$ & $-10.62$ & $-12.18 \pm 0.14$ & $2.2 \pm 3.8$ \\
Fornax & $149$ & $-11.20$ & $-11.35 \pm 0.17$ & $2.2 \pm 1.1$ \\
Leo I & $258$ & $-11.68$ & $-11.06 \pm 0.18$ & $2.6 \pm 1.5$ \\
Leo II & $236$ & $-11.60$ & $-11.67 \pm 0.26$ & $2.8 \pm 2.3$ \\
Sculptor & $86$ & $-10.72$ & $-11.48 \pm 0.26$ & $2.1 \pm 3.0$ \\
Sextans & $89$ & $-10.75$ & $-12.96 \pm 0.24$ & $1.7 \pm 1.1$ \\
Ursa Minor & $78$ & $-10.64$ & $-12.60 \pm 0.25$ & $2.0 \pm 2.2$ \\
Bootes I & $64$ & $-10.47$ & $-13.47 \pm 0.14$ & $1.5 \pm 1.0$ \\
Bootes II & $40$ & $-10.06$ & $-13.40 \pm 0.53$ & $0 \pm 220$ \\
Canes V. I & $216$ & $-11.52$ & $-13.40 \pm 0.14$ & $2.40 \pm 0.41$ \\
Canes V. II & $161$ & $-11.27$ & $-12.76 \pm 0.14$ & $2.7 \pm 3.2$ \\
Coma B. & $45$ & $-10.16$ & $-13.19 \pm 0.27$ & $1.4 \pm 3.6$ \\
Hercules & $128$ & $-11.07$ & $-12.92 \pm 0.22$ & $2.0 \pm 1.6$ \\
Horologium I & $79$ & $-10.65$ & $-12.78 \pm 0.17$ & $3 \pm 49$ \\
Hydra II & $131$ & $-11.09$ & $-12.88 \pm 0.28$ & $3 \pm 16$ \\
Leo IV & $155$ & $-11.24$ & $-13.30 \pm 0.30$ & $2.1 \pm 3.1$ \\
Leo V & $176$ & $-11.35$ & $-13.03 \pm 0.38$ & $2.0 \pm 4.3$ \\
Leo T & $422$ & $-12.11$ & $-11.70 \pm 0.14$ & $4.0 \pm 6.3$ \\
Pisces II & $195$ & $-11.44$ & $-12.91 \pm 0.29$ & $4 \pm 27$ \\
Reticulum II & $31$ & $-9.84$ & $-13.10 \pm 0.13$ & $1 \pm 10$ \\
Segue I & $28$ & $-9.75$ & $-13.22 \pm 0.45$ & $1 \pm 21$ \\
Segue II & $41$ & $-10.08$ & $-13.17 \pm 0.20$ & $1 \pm 11$ \\
Tucana & $883$ & $-12.75$ & $-11.98 \pm 0.21$ & $20 \pm 140$ \\
Ursa Major I & $101$ & $-10.86$ & $-13.27 \pm 0.27$ & $2.6 \pm 4.3$ \\
Ursa Major II & $38$ & $-10.02$ & $-13.49 \pm 0.23$ & $1.2 \pm 6.8$ \\
Willman I & $43$ & $-10.12$ & $-12.63 \pm 0.36$ & $2 \pm 28$ \\
				
			\end{tabular}
			\tablefoot{Columns from left to right: Name of the galaxy, Distance $D_\text{host}$ Milky Way and galaxy, Milky Way field $g_\text{host}$ on galaxy, Newtonian field $g_\text{N}$ of galaxy, $p$-parameter obtained from $g_\text{host}+g_\text{N}$. $D_\text{host}$, $g_\text{N}$ taken from \cite{2017ApJ...836..152L}. $g_\text{host}$ calculated from Eq. \eqref{eq:ghost}, $p$ from Eq. \eqref{eq:pgtot}.}
		\end{table}
	
		\begin{table}[h!]
		\section{External field Andromeda}

			\caption{Relevant data on testing for the effect of the Andromeda gravitational field on Andromeda dSphs.}
			\label{tab:andext}
			\begin{tabular}{l|r|r|r|r}
				Galaxy &{$\frac{D_\text{host}}{ \unit{\kilo\parsec}}$}&{$\log_{10}(\frac{g_\text{host}}{\unit{\metre\per\second\squared}})$}& {$\log_{10}(\frac{g_\text{N}}{\unit{\metre\per\second\squared}})$} & {$p$}\\
				\midrule
				NGC 147 & $118$ & $-10.70$ & $-10.67 \pm 0.12$ & $1.9 \pm 2.3$ \\
NGC 185 & $181$ & $-11.07$ & $-10.52 \pm 0.12$ & $7 \pm 64$ \\
NGC 205 & $46$ & $-9.88$ & $-9.88 \pm 0.19$ & $3 \pm 31$ \\
And I & $68$ & $-10.22$ & $-12.06 \pm 0.15$ & $1.3 \pm 1.8$ \\
A. II & $195$ & $-11.13$ & $-12.27 \pm 0.16$ & $1.9 \pm 0.7$ \\
A. III & $86$ & $-10.42$ & $-12.23 \pm 0.18$ & $1.4 \pm 1.8$ \\
A. V & $113$ & $-10.66$ & $-12.43 \pm 0.19$ & $2.2 \pm 3.0$ \\
A. VI & $268$ & $-11.41$ & $-11.88 \pm 0.16$ & $3.2 \pm 2.8$ \\
A. VII & $217$ & $-11.23$ & $-11.85 \pm 0.17$ & $2.4 \pm 1.0$ \\
A. IX & $182$ & $-11.07$ & $-13.42 \pm 0.45$ & $2.5 \pm 2.8$ \\
A. X & $92$ & $-10.48$ & $-12.72 \pm 0.43$ & $1.4 \pm 1.4$ \\
A. XI & $102$ & $-10.57$ & $-12.80 \pm 0.49$ & $2 \pm 11$ \\
A. XII & $181$ & $-11.07$ & $-13.58 \pm 0.51$ & $1.8 \pm 3.8$ \\
A. XIII & $115$ & $-10.68$ & $-13.12 \pm 0.50$ & $1.8 \pm 3.7$ \\
A. XIV & $161$ & $-10.97$ & $-12.76 \pm 0.48$ & $1.8 \pm 1.8$ \\
A. XV & $174$ & $-11.04$ & $-12.11 \pm 0.17$ & $1.8 \pm 1.7$ \\
A. XVI & $319$ & $-11.56$ & $-11.76 \pm 0.17$ & $2.8 \pm 2.8$ \\
A. XVII & $67$ & $-10.21$ & $-12.47 \pm 0.21$ & $1.3 \pm 5.0$ \\
A. XIX & $116$ & $-10.68$ & $-13.84 \pm 0.26$ & $1.38 \pm 0.76$ \\
A. XX & $128$ & $-10.77$ & $-12.67 \pm 0.45$ & $3 \pm 20$ \\
A. XXI & $135$ & $-10.82$ & $-13.01 \pm 0.28$ & $1.50 \pm 0.76$ \\
A. XXII & $275$ & $-11.43$ & $-12.89 \pm 0.41$ & $2.0 \pm 2.5$ \\
A. XXIII & $127$ & $-10.76$ & $-12.88 \pm 0.24$ & $1.60 \pm 0.68$ \\
A. XXIV & $169$ & $-11.01$ & $-13.14 \pm 0.24$ & $1.9 \pm 6.3$ \\
A. XXV & $90$ & $-10.46$ & $-12.85 \pm 0.24$ & $1.27 \pm 0.87$ \\
A. XXVI & $103$ & $-10.58$ & $-12.90 \pm 0.25$ & $2.0 \pm 7.8$ \\
A. XXVII & $482$ & $-11.92$ & $-13.07 \pm 0.40$ & $4.4 \pm 8.9$ \\
A. XXVIII & $384$ & $-11.72$ & $-12.37 \pm 0.20$ & $2.5 \pm 2.1$ \\
A. XXIX & $198$ & $-11.15$ & $-12.68 \pm 0.18$ & $2.1 \pm 1.4$ \\
A. XXX & $145$ & $-10.88$ & $-12.63 \pm 0.17$ & $3 \pm 31$ \\
A. XXXI & $262$ & $-11.39$ & $-12.43 \pm 0.32$ & $2.24 \pm 0.72$ \\
A. XXXII & $140$ & $-10.85$ & $-12.31 \pm 0.37$ & $1.66 \pm 0.73$ \\
Cetus & $688$ & $-12.23$ & $-12.22 \pm 0.16$ & $2.74 \pm 0.84$ \\
Perseus I & $351$ & $-11.65$ & $-12.00 \pm 0.45$ & $2.1 \pm 4.2$ \\
Pisces I & $268$ & $-11.41$ & $-12.37 \pm 0.14$ & $2.4 \pm 5.2$ \\
			\end{tabular}
			\tablefoot{Columns from left to right: name of galaxy, distance $D_\text{host}$ between Andromeda and galaxy, Andromeda gravitational field $g_\text{host}$ exerted on galaxy, Newtonian field $g_\text{N}$ of galaxy, $p$-parameter from p-Laplacian inferred from $g_\text{host}+g_\text{N}$. $D_\text{host}$ and $g_\text{N}$ taken from \cite{2017ApJ...836..152L}. $g_\text{host}$ inferred from Eq. \eqref{eq:ghost}. $p$ calculated from Eq. \eqref{eq:pgtot}.}
		\end{table}
	
		\begin{figure}[h!]
			\section{External field additional plots}
			\centering
			\begin{subfigure}{0.5\textwidth}
				\includegraphics[width=\textwidth]{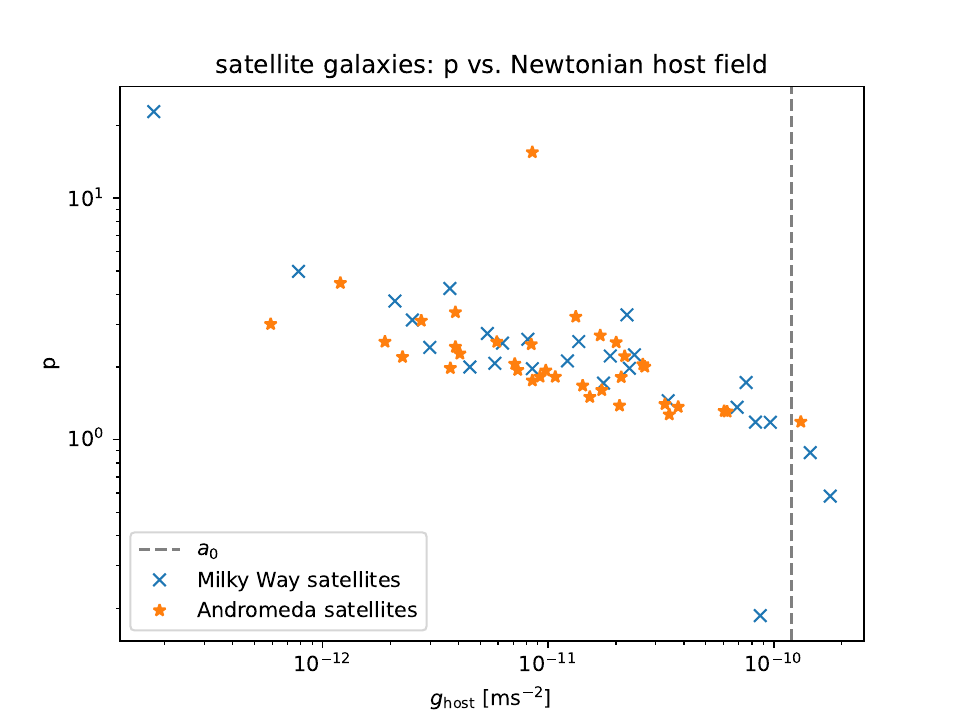}
				\caption{$p$ vs. $g_\text{host}$} 
				\label{fig:accghost}
			\end{subfigure}
			\begin{subfigure}{0.5\textwidth}
				\includegraphics[width=\textwidth]{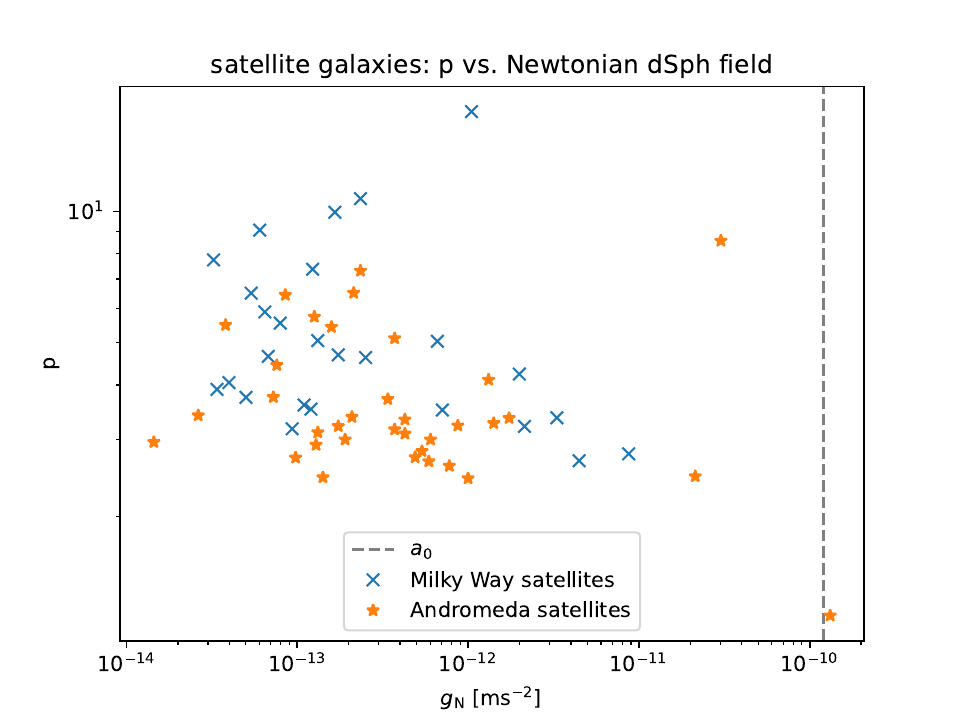}
				\caption{$p$ vs. $g_\text{N}$}
				\label{fig:accgN}
			\end{subfigure}
			\caption{Relation between $p$-parameter and external field for Milky Way (blue) and Andromeda (orange) satellite dSphs. Uncertainties are not shown to keep the plot legible. The dashed line describes Milgrom's constant $a_0$. The external field is Newtonian and consists of either the field by the host galaxy \ref{fig:accghost} or the internal field of the satellite \ref{fig:accgN}. The dependence of $p$ on the host field is more obvious.}	
		\end{figure}
		\twocolumn
		\section{Derivation of velocity dispersion independently of density profile}
		\label{sec:app2}
		In the following, the velocity dispersion for arbitrary $p$ neglecting the EFE is derived. Furthermore, it is shown, that for arbitrary $p$, numerical values for the velocity dispersion can only be calculated with specifying a density profile. \citet{Milgrom_1994} derived the velocity dispersion for $p=3$. This derivation is analogous to his.
		Milgrom considered the quantity $Q$, integrating over phase-space:
		\begin{equation}
			Q := \sum_k \int m_k f_k(\vec{r},\vec{v},t)\, \vec{r} \cdot \vec{v}\, \text{d}^3r \text{d}^3v\;.
		\end{equation}
		$m_k$ are the masses of arbitrary particles (e.g. stars). $f_k(\vec{r},\vec{v},t)$ is a phase-space-distribution function. 
		Milgrom showed that its time derivative $\dot{Q}$ can be written as:
		\begin{equation}
			\dot{Q} = M \langle v^2 \rangle (t) - \int \rho(\vec{r},t)\, \vec{r} \cdot \nabla \Phi \, \text{d}^3r\;.
			\label{eq:qder}
		\end{equation}
		Here, the finite total mass $M$ of the system and the momentary, mass-weighted, 3-D, mean-square velocity $\langle v^2 \rangle (t)$ is introduced.
		The GPE is used:
		\begin{align}
			\nabla \cdot \left((\mu(x)) \nabla \Phi\right) &= 4\pi G \rho\;, \label{eq:genpoiseq} \\ 
			\mu(x) &:= x^{p-2}\;, \label{eq:mu}\\
			x&:=\frac{\left|\nabla\Phi\right|}{a_0}\;.
		\end{align}
		As opposed to Milgrom, $\mu(x)$ is here defined in terms of a fixed $p>1$.
		Milgrom inserted the density distribution $\rho$ described by Eq. \eqref{eq:genpoiseq} into Eq. \eqref{eq:qder}. Furthermore, $F(y)$ is introduced, such that:
		\begin{equation}
			\mu(x) = \frac{\text{d}F(y)}{\text{d}y},\; y=x^2\;.
		\end{equation}
		Since in the case for arbitrary $p$, $\mu(x)$ is defined by Eq. \eqref{eq:mu}:
		\begin{equation}
			F(x^2) = \frac{2}{p}x^p \;.
		\end{equation}
		Milgrom showed, that $\dot{Q}$ can be written as:
		\begin{align}
			\dot{Q} &= M \langle v^2 \rangle (t) + I_1 + I_2\;, \label{eq:qdot}\\
			I_1 &:= - \frac{1}{4\pi G} \int \mu(x) \vec{r}\cdot\nabla\Phi\,\nabla\Phi\cdot\text{d}\vec{s}\;,\label{eq:i1}\\
			I_2 &:= I_3 + I_4\;, \\
			I_3 &:= \frac{a_0^2}{8\pi G} \int F(x^2) \vec{r} \cdot \text{d}\vec{s}\;, \label{eq:i3}\\
			I_4 &:= - \frac{a_0^2}{4\pi G} \int \frac{3}{2} F(x^2) - \mu(x)x^2\, \text{d}^3r \;. \label{eq:i4}
		\end{align}
		Milgrom solved Eq. \eqref{eq:qdot} for $\langle v^2 \rangle (t)$ and took the long-time average $\overline{\langle v^2 \rangle}$. Doing so for arbitrary $p$:
		\begin{equation}
			\overline{\langle v^2 \rangle} = - \frac{1}{M} (I_1+I_3+I_4)
		\end{equation}
		Assume the density distribution to have a sphere with radius $R$ as compact support and $\nabla\Phi$ at $R$ for arbitrary $p$ to behave like the gravitational acceleration produced by a point mass:
		\begin{equation}
			\nabla\Phi = \sqrt[p-1]{\frac{GMa_0^{p-2}}{R^2}}\, \vec{\text{e}}_r\;.
		\end{equation}
		Equation \ref{eq:i1} is evaluated at $R$:
		\begin{align}
			I_1 &= - \frac{1}{4\pi G} \left(\frac{\left|\nabla\Phi\right|}{a_0}\right)^{p-2} R \left|\nabla\Phi\right| 4\pi R^2 \left|\nabla\Phi\right|\\
			&= - \frac{R^3}{G} a_0^{2-p} \left|\nabla\Phi\right|^p \\
			&= - M\sqrt[p-1]{GMR^{p-3}a^{p-2}_0} \;.
		\end{align}
		Equation \ref{eq:i3} is evaluated at $R$:
		\begin{align}
			I_3 &= \frac{a_0^2}{8\pi G} \frac{2}{p}\left(\frac{\left|\nabla\Phi\right|}{a_0}\right)^p 4 \pi R^3 \\
			&= \frac{1}{p} \frac{R^3}{G} a_0^{2-p} \left|\nabla\Phi\right|^p \\
			&= \frac{1}{p} M\sqrt[p-1]{GMR^{p-3}a^{p-2}_0} \; .
		\end{align}
		Equation \ref{eq:i4} is evaluated at $R$:
		\begin{align}
			I_4 &= - \frac{a_0^2}{4\pi G} \int \left(\frac{3}{2} \frac{2}{p}x^p - x^{p-2} x^2 \right) \, \text{d}^3r \\
			&= - \frac{a_0^2}{4\pi G} \int \frac{3-p}{p}x^p \, \text{d}^3r \\
			&= - \frac{a_0^2}{4\pi G} \frac{3-p}{p} \int_0^R 4 \pi r^2 \left(\frac{a(r)}{a_0}\right)^p \, \text{d}r \\
			&= - \frac{a_0^2}{G} \frac{3-p}{p} \int_0^R r^2 \left(\frac{a(r)}{a_0}\right)^p \, \text{d}r \\
			&= - \frac{1}{G} \frac{3-p}{p} \int_0^R r^2 \left(\frac{a(r)}{a_0}\right)^{p-2} a^2(r) \, \text{d}r \\
			&= - \frac{1}{G} \frac{3-p}{p} \int_0^R GM(r) a(r) \, \text{d}r \\
			&=  \frac{p-3}{p} \int_0^R M(r) a(r) \, \text{d}r \;.
		\end{align}
		$I_4$ depends on the density distribution. It vanishes only for $p=3$ (Milgromian case). This explains, why the velocity dispersion is independent of the density distribution only for Milgromian gravity.
		
	\end{appendix}
\end{document}